\documentclass[aps,preprint]{revtex4}
\usepackage{amssymb,epsf}
\usepackage{amsmath}
\usepackage{graphicx}

\begin{document}

\title{Thermodynamic Instability of Black Holes of Third Order
Lovelock Gravity}
\author{M. H. Dehghani$^{1,2}$\footnote{email address:
mhd@shirazu.ac.ir} R. Pourhasan$^{1}$}
\affiliation{$^1$ Physics Department and Biruni Observatory, College of Sciences, Shiraz University, Shiraz 71454, Iran\\
        $^2$ Research Institute for Astrophysics and Astronomy of Maragha (RIAAM), Maragha,
        Iran}

\begin{abstract}
In this paper, we compute the mass and the temperature of the uncharged
black holes of third order Lovelock gravity and compute the
entropy through the use of first law of thermodynamics.
We perform a stability analysis by studying the curves of temperature versus
the mass parameter, and find that there exists an intermediate
thermodynamically unstable phase
for black holes with hyperbolic horizon. The existence of
this unstable phase for the uncharged topological black holes of
third order Lovelock gravity does not occur in the lower
order Lovelock gravity. We also perform a stability analysis
for a spherical, 7-dimensional black hole of Lovelock gravity and find that while
these kinds of black holes for small values of Lovelock coefficients
have an intermediate unstable phase, they are stable for large values
of Lovelock coefficients. We also find that there exists
an intermediate unstable phase for these black holes in higher dimensions.
This stability analysis shows that the thermodynamic stability of
black holes with curved horizons is not a robust feature of all the generalized theories
of gravity.
\end{abstract}

\pacs{04.50.-h, 04.70.Bw, 04.70.Dy, 04.70.-s}
\maketitle

\section{Introduction}

Thermodynamics of black holes in anti-de Sitter (AdS) spaces have been the subject
of wide variety of researches in recent years. This is due to the
fact that, in parallel with the development of AdS/CFT
correspondence \cite{AdSCFT}, black holes in AdS spaces are known
to play an important role in dual field theory. With the
AdS/CFT correspondence, one can
gain some insights into thermodynamic properties and phase
structures of strong 't Hooft coupling CFTs by studying
thermodynamics of AdS black holes. In the context of horizon
topology, asymptotically AdS black holes are
quite different from black holes in flat or dS spaces. In
asymptotically flat or dS spaces, the horizon topology of a four
dimensional black hole must be a round sphere $S^{2}$
\cite{FrJa}, while in AdS spaces it is possible to have
black holes with zero or negative constant curvature horizon too.
These black holes are referred to as topological black holes in
the literature. Due to the different horizon structures, the
associated thermodynamic properties of topological black holes
are rather different from the spherically symmetric black holes
\cite{Lemos}. While the Schwarzschild black hole in AdS space is thermodynamically stable for large mass, it becomes unstable for small mass. That is, there is a phase
transition (namely Hawking-Page phase transition) between the
high temperature black hole phase and low temperature thermal AdS
space \cite{HawPage}. It has been explained by Witten
\cite{EWitten} that the Hawking-Page phase transition of
Schwarzschild black holes in AdS spaces can be identified with
confinement/deconfinement transition of the Yang-Mills theory in
the AdS/CFT correspondence. However, it is interesting to note
that for the AdS black holes with zero or negative constant
curvature horizon the Hawking-Page phase transition does not
appear and these black holes are always locally stable
\cite{birmingham} (see also \cite{Brill}).

Now, the question which arises is that whether the stability of
black hole is a robust feature of all generally
covariant theories of gravity or is peculiar to Einstein
gravity. Among gravity theories, the so-called Lovelock gravity  \cite{lovelock} has some special features.
For example, the resulting field equations contain no more
than second derivatives of the metric and it has been proven to
be ghost-free when expanding about the flat space, evading any
problem with unitarity. In this paper, we investigate the
stability of uncharged black holes of third order
Lovelock gravity with hyperbolic horizon. It is, now, known that the
topological black holes of third order Lovelock gravity with
zero curvature horizon is thermodynamically stable \cite{deh}.
Indeed, all the thermodynamic and conserved quantities of the black holes
with flat horizon do not depend on the Lovelock coefficients, and therefore
these black holes are stable as the Einstein's black hole
with flat horizon. This phase behavior of black holes with flat horizon
is also commensurate with the fact that
there is no Hawking-Page transition for a black object
whose horizon is diffeomorphic to $\Bbb{R}^{p}$ \cite{EWitten}.
Also, as in the case of Einstein gravity \cite{birmingham},
the black holes of Gauss-Bonnet gravity with hyperbolic horizon
is stable \cite{Cai1}. These facts bring in the idea that the
Lovelock terms may have no effect on the stability of topological
black holes. But, one of us has shown that an asymptotically
flat uncharged black hole of third order Lovelock gravity may have two horizons \cite{Sha},
a fact that does not happen in lower order Lovelock gravity.
This persuades us to investigate the effects of third order Lovelock
term on the stability phase structure of the black holes with curved horizon.
We show that the hyperbolic uncharged black holes of
third order Lovelock gravity have an intermediate
unstable phase in contrast to the uncharged topological black
holes of Einstein gravity or Gauss-Bonnet gravity.
We also, investigate the effects of third order Lovelock term on the stability of a spherical
black hole of third order Lovelock gravity.

The outline of this paper is as follows. We give a brief review
of the Hamiltonian formulation of Lovelock action in Sec.
(\ref{HF}). In Sec. (\ref {3rdL}) we obtain the vacuum solutions
of third order Lovelock gravity by using the Hamiltonian form of the
action and discuss the thermodynamics of the solutions.
We investigate the stability
of the uncharged black holes  with curved horizon in Sec. (\ref{stunsol}).
Finally, we finish our paper with some concluding remarks.

\section{Hamiltonian Formulation}
\label{HF}
The most fundamental assumption in standard general relativity is
the requirement that the field equations be generally covariant
and contain at most second order derivatives of the metric. Based
on this principle, the most general classical theory of
gravitation in an $(n+1)$-dimensional manifold $\mathcal{M}$ with
the metric $g_{\mu \nu }$\ is Lovelock gravity \cite{lovelock},
for which the gravitational action may be written as
\begin{equation}
I_{G}=\frac{1}{16\pi}\int d^{n+1}x\sqrt{-g}\sum_{p=0}^{[n/2]}\alpha _{p}%
\mathcal{L}_{p},  \label{lovact}
\end{equation}
where $[n/2]$ denotes the integer part of $n/2$, $\alpha _{p}$'s
are Lovelock coefficients and
\begin{equation}
\mathcal{L}_{p}=\frac{1}{2^{p}}\delta _{\rho _{1}\sigma
_{1}\cdots \rho _{p}\sigma _{p}}^{\mu _{1}\nu _{1}\cdots \mu
_{p}\nu _{p}}R_{\mu _{1}\nu
_{1}}^{\phantom{\mu_1\nu_1}{\rho_1\sigma_1}}\cdots R_{\mu _{p}\nu _{p}}^{%
\phantom{\mu_k \nu_k}{\rho_p \sigma_p}}  \label{Lp}
\end{equation}
is the Euler density of a $2p$-dimensional manifold. In Eq. (\ref{Lp}) $%
\delta _{\rho _{1}\sigma _{1}\cdots \rho _{p}\sigma _{p}}^{\mu
_{1}\nu _{1}\cdots \mu _{p}\nu _{p}}$ is a totally
antisymmetric product of Kronecker delta and $R_{\mu \nu
}^{\phantom{\mu\nu}{\rho\sigma}}$ is the Riemann tensor of the
Manifold $\mathcal{M}$. In $n + 1$ dimensions, all terms for which $p >\lbrack n/2$ are total derivatives and therefore only the terms for which $%
p\leq \lbrack n/2]$ contribute to the field equations.

In order to simplify the equations of motion, it is more
convenient to use the Hamiltonian formulation. This formulation
requires a breakup of spacetime into space and time which yields
some insights into the nature of the dynamics of general
relativity. Indeed, in this approach the dynamical variable is
the spatial metric $h_{ij}$ rather than spacetime metric $g^{\mu
\nu }$, where $h_{ij}$ is the induced metric on the spacelike hypersurface $%
\Sigma _{t}$ of the spacetime manifold $\mathcal{M}$. In this approach, the canonical coordinates are
the spatial components of the metric $g_{ij}$, and their
conjugate momenta are \cite{zanelli}
\begin{eqnarray}
\pi _{\phantom{i}{j}}^{i}
&=&-\frac{1}{4}\sqrt{-g}\sum_{p=0}^{n}\frac{\alpha
_{p}}{2^{p}}\sum_{s=0}^{p-1}\frac{(-4)^{p-s}}{s![2(p-s)-1]!!}\delta
_{\lbrack j_{1}\ldots j_{2p-1}j}^{i_{1}\ldots i_{2p-1}i}  \nonumber \\
&&\times \hat{R}_{\phantom{j_1j_2}{i_1i_2}}^{j_{1}j_{2}}\cdots \hat{R}_{%
\phantom{j_{2s-1}j_{2s}}{i_{2s-1}i_{2s}}}^{j_{2s-1}j_{2s}}K_{%
\phantom{j_{2s+1}}{i_{2s+1}}}^{j_{2s+1}}\cdots K_{%
\phantom{j_{2p-1}}{i_{2p-1}}}^{j_{2p-1}}\,,
\end{eqnarray}
where $K_{\phantom{i}{j}}^{i}$ is the extrinsic curvature of the
hypersurface $\Sigma _{t}$ given as:
\begin{equation}
K_{ij}=\frac{1}{2}N^{-1}(\dot{h}_{ij}-D_j N_{i}-D_i N_{j})\,,
\label{ext}
\end{equation}
and $\hat{R}_{ijkl}$ are the components of the intrinsic
curvature tensor
of the boundary $\Sigma _{t}$. In Eq. (\ref{ext}), $N=(-g^{00})^{-1/2}$ and $%
N^{i}=h^{ij}g_{0\mu }$ are the `\textit{lapse function}' and the `\textit{%
shift vectors}' in the standard ADM (Arnowitt-Deser-Misner) decomposition of spacetime,
and $D_{i}$ denotes the covariant derivative associated with
$h_{ij}$. The time components $g_{0\mu }$ are Lagrange
multipliers associated with the generators of surface deformation
\begin{equation}
\mathcal{H}=-\sqrt{h}\sum_{p}\alpha _{p}\frac{1}{2^{p}}\delta
_{j_{1}\cdots j_{2p}}^{i_{1}\cdots
i_{2p}}R_{\phantom{j_1j_2}{i_1i_2}}^{j_{1}j_{2}}\cdots
R_{\phantom{j_{2p-1}j_{2p}}{i_{2p-1}i_{2p}}}^{j_{2p-1}j_{2p}},
\label{HH}
\end{equation}
and $\mathcal{H}_{i}=-2\pi _{\;i|j}^{j}$. In Eq. (\ref{GC}), $R_{%
\phantom{ij}{kl}}^{ij}$ are the spatial components of spacetime
curvature tensor, which depend on the velocities through the
Gauss--Codacci equation
\begin{equation}
R_{ijkl}=\hat{R}_{ijkl}+K_{ik}K_{jl}-K_{il}K_{jk}\,.  \label{GC}
\end{equation}
In this formalism, the action (\ref{lovact}) becomes
\begin{equation}
I_{G}=\frac{1}{16\pi}\int dtd^{n}x(\pi ^{ij}\dot{h}_{ij}-N\mathcal{H}-N^{i}%
\mathcal{H}_{i})+B,  \label{B1}
\end{equation}
where $B$ stands for a surface term.

\section{Third Order Lovelock Black Holes in AdS Space}
\label{3rdL}
The metric of an ($n+1$)-dimensional static spherically symmetric
spacetime may be written as
\begin{equation}
ds^{2}=-f(r)dt^{2}+\frac{dr^{2}}{f(r)}+r^{2}d\Sigma
_{k,n-1}^{2}, \label{metric}
\end{equation}
where $d\Sigma _{k,n-1}^{2}$ represents the metric of an
$(n-1)$-dimensional hypersurface with constant curvature
$(n-1)(n-2)k$ and volume $V_{n-1}$.

We consider the third order Lovelock gravity, and therefore
we restrict ourselves to the first four terms of the
Hamiltonian form of the action (\ref{B1}). Using the Gauss-Codacci relation (\ref{GC}) and the fact that the extrinsic curvature $K_{ij}$ is zero for the metric in Eq. (\ref{metric}), the generator of surface deformation becomes $\mathcal{H}=\sum \alpha_p \mathcal{L}_p$, and consequently the action (7) (with the first four terms) reduces to
\begin{equation}
I_{G}=\frac{1}{16\pi}\int dtd^{n}xN\sqrt{h}[-2\Lambda +\mathcal{L}%
_{1}+\alpha _{2}\mathcal{L}_{2}+\alpha _{3}\mathcal{L}_{3}]+B,
\label{IG}
\end{equation}
where $\Lambda =-n(n-1)/2l^{2}$ is the cosmological constant for
AdS solutions, and $\alpha _{2}$ and $\alpha _{3}$ are Gauss-Bonnet
and third order
Lovelock coefficients with dimensions $(\mathrm{length})^{2}$ and $(\mathrm{%
length})^{4}$, respectively, which are assumed to be positive. In Eq. (\ref{IG}), $\mathcal{L}%
_{1}={R}$ is just the Einstein-Hilbert Lagrangian, $\mathcal{L}_{2}=
R_{ijkl}{R}^{ijkl}-4{R}_{ij}{R}^{ij}+{R}^{2}$ is
the second order Lovelock (Gauss-Bonnet) Lagrangian, and
\begin{eqnarray}
\mathcal{L}_{3}\hspace{-0.2cm} &=&\hspace{-0.2cm}2{R}^{ijkl}{R}%
_{klmn}R_{\phantom{mn}{ij}}^{mn}+8R_{\phantom{ij}{km}}^{ij}{R%
}_{\phantom{kl}{jn}}^{kl}R_{\phantom{mn}{il}}^{mn}+24R^{ijkl}%
R_{kljm}R_{\phantom{m}{i}}^{m}  \nonumber \\
&&\hspace{-0.1cm}+3RR^{ijkl}R_{klij}+24R^{ikjl}{R%
}_{ji}R_{lk}+16R^{ij}R_{jk}R_{\phantom{k}{i}}^{k}-12%
RR^{ij}R_{ji}+R^{3}
\end{eqnarray}
is the third order Lovelock Lagrangian.
Defining the dimensionless parameters $a$ and $b$ as
\begin{equation}
a=\frac{(n-2)(n-3)}{l^{2}}\alpha _{2},\hspace{.5cm}
b=\frac{72}{l^{4}} \left(_{\phantom{n}{4}}^{n-2}\right)\alpha_{3}\,,
\end{equation}
the action (\ref{IG}) reduces to:
\begin{equation}
I_{G}=\frac{(n-1)V_{n-1}}{16\pi}\int dtdr\left[ \frac{r^{n}}{l^{2}}%
+r^{n}\psi (1+l^{2}a\psi +\frac{l^{4}}{3}b\psi ^{2})\right]
^{\prime }+B, \label{Igfin}
\end{equation}
where a prime denotes the derivative with respect to $r$ and $\psi
=r^{-2}[k-f(r)]$. The surface term is $B=-(t_2-t_1)M+B_0$, where
$B_0$ is an arbitrary constant and $M$ is the conserved charge associated to the
time displacement \cite{zanelli}.
Varying the action (\ref{Igfin}) with respect to $f$, one obtains
the equation of motion as
\begin{equation}
\left[\frac{r^{n}}{l^{2}}+r^{n}\psi (1+l^{2}a\psi
+\frac{l^{4}}{3}b\psi ^{2})\right] ^{\prime }=0. \label{Efr}
\end{equation}
The only real solution of Eq. (\ref{Efr}) is
\begin{eqnarray}
f(r)&=&k+\frac{ar^{2}}{bl^{2}}\left[ 1+\left( \sqrt{\Gamma +J^{2}(r)}%
+J(r)\right) ^{1/3}-\Gamma ^{1/3}\left( \sqrt{\Gamma
+J^{2}(r)}+J(r)\right) ^{-1/3}\right] \nonumber\\
&=&k+\frac{ar^{2}}{bl^{2}}\left[ 1+\left( \sqrt{\Gamma +J^{2}(r)}%
+J(r)\right) ^{1/3}-\left( \sqrt{\Gamma
+J^{2}(r)}-J(r)\right) ^{1/3}\right], \label{Fr}
\end{eqnarray}
where
\begin{eqnarray}
&&\Gamma =\left( \frac{b}{a^{2}}-1\right) ^{3}\,,  \label{gamma} \\
&&J(r)=1-\frac{3b}{2a^{2}}+\frac{3b^{2}}{2a^{3}}K(r)\,,  \label{jr} \\
&&K(r)=1-\frac{ml^{2}}{r^{n}}\,.
\end{eqnarray}
and $m$ is an integration constant. The metric function $f(r)$ is
real everywhere provided
\begin{equation}
9b^{2}+(4-18a)b+(12a^{3}-3a^{2})\geq 0.  \label{conditionl}
\end{equation}
The above condition (\ref{conditionl}) is satisfied for the case
of $a\geq 1/3$, while for the case of $a<1/3$, it is satisfied
provided $b>b^{(+)}$ or $b<b^{(-)}$, where $b^{(+)}$ and
$b^{(-)}$ are the larger and smaller real roots of Eq. (\ref{conditionl}),
respectively.

The ADM mass of black hole can be
obtained by using the behavior of the metric at large $r$. It is
easy to show that the mass of the black hole per unit volume,
$V_{n-1}$, is
\begin{equation}
M=(n-1)m/16\pi,  \label{Mass}
\end{equation}
where the mass parameter $m$ in
terms of the real roots of $f(r_h)=0$ is:
\begin{equation}
m(r_{h})=l^{-2}r_{h}^{n}+kr_{h}^{n-2}+k^{2}al^{2}r_{h}^{n-4}+\frac{kbl^{4}}{3%
}r_{h}^{n-6}.  \label{bmass}
\end{equation}
The Hawking temperature
of the black holes can be obtained by requiring the
absence of conical singularity at the event horizon in the
Euclidean sector of the black hole solution as:
\begin{equation}
T=\frac{f^{\prime }(r_{+})}{4\pi }=\frac{%
3nr_{+}^{6}+3k(n-2)l^{2}r_{+}^{4}+3k^{2}(n-4)al^{4}r_{+}^{2}+k(n-6)bl^{6}}{%
12\pi l^{2}r_{+}(r_{+}^{4}+2kal^{2}r_{+}^{2}+k^{2}bl^{4})},
\label{btemp}
\end{equation}
where $r_{+}$ is the radius of event horizon. Clearly, the
temperature is always positive for $k=0$ and $k=1$ cases, and
therefore there is no extreme black holes. However, for the case
of $k=-1$, the black hole
solutions may present an extreme black hole with horizon radius $r_{\mathrm{%
ext}}$, where $r_{\mathrm{ext}}$ is one of the real roots of $T=0$.

Usually, the entropy of black holes satisfies the so-called area law
of entropy which states that the black hole entropy equals
one-quarter of the horizon area \cite{Bek}. However, in higher
derivative gravity the area law of entropy is not satisfied in
general \cite{fails}. A simple method of finding the entropy is
through the use of first law of thermodynamics, $dM=TdS$
\cite{Wald}. When the mass parameter is nonnegative, the horizon
radius starts from zero. Integrating the first law
\begin{equation}
S=\int_{0}^{r_{+}}T^{-1}\left( \frac{\partial M}{\partial
r_{+}}\right) dr_{+},  \label{bent}
\end{equation}
one obtains the entropy per unit volume $V_{n-1}$ as:
\begin{equation}
S=\frac{r_{+}^{n-1}}{4}\left( 1+\frac{2k(n-1)al^{2}}{(n-3)r_{+}^{2}}+\frac{%
k^{2}(n-1)bl^{4}}{(n-5)r_{+}^{4}}\right) ,  \label{Ent}
\end{equation}
which reduces to the area law of entropy for $a=b=0$. Although
for the case of $k=-1$, the mass parameter $m$ may be negative
and the black hole horizon can not shrink to zero, the entropy
given by Eq. (\ref{Ent}) is applicable, since it reduces to the
area law of entropy for Einstein gravity.
One may find that the entropy per unit volume obeys the law of
the entropy of asymptotically flat black holes of $p$th-order
Lovelock gravity \cite{Myers}
\begin{equation}
S=\frac{1}{4}\sum_{k=1}^{p}k\alpha _{k}\int d^{n-1}x\sqrt{\tilde{g}}\tilde{%
\mathcal{L}}_{k-1},  \label{Enta}
\end{equation}
where the integration is done on the $(n-1)$-dimensional spacelike
hypersurface of Killing horizon, $\tilde{g}_{\mu \nu }$ is the
induced
metric on it, $\tilde{g}$ is the determinant of $\tilde{g}_{\mu \nu }$ and $%
\tilde{\mathcal{L}}_{k}$ is the $k$th order Lovelock Lagrangian of $\tilde{g}%
_{\mu \nu }$.

All the thermodynamic quantities
obtained in this section reduce to those of Einstein gravity
given in \cite{birmingham} for $a=b=0$.

\section{Stability of the Uncharged Black Holes}
\label{stunsol}
In this section, we consider the special case of $b=a^{2}$ for which the
metric function $f(r)$ becomes
\begin{equation}
f(r)=k+\frac{r^{2}}{al^{2}}\left[ 1-\left( 1-3a+\frac{3aml^{2}}{r^{n}}%
\right) ^{1/3}\right] ,  \label{frs}
\end{equation}
which is real everywhere. The thermodynamic quantities may be
obtained by substituting $b=a^{2}$ in those obtained in the
previous section. In the uncharged case, the positivity of the heat
capacity $C=\partial M/\partial T$ is sufficient to ensure the local
stability, and therefore the plot of $T $ versus $m$ gives all
the information about thermodynamic stability. For $k=0$, the mass,
the temperature and the entropy do not depend on the Lovelock coefficients, as one may see
from Eqs. (\ref{Mass})-(\ref{Ent}), and the black hole with flat horizon
is stable \cite{deh}. The stability of black holes with curved horizon, which is the main goal of
this paper, will be discussed in the rest of the paper.

\subsection{Seven-dimensional hyperbolic black holes}
First, we study the stability of 7-dimensional black holes of
third order Lovelock gravity, which is the most general
solution of gravity based on the principle of the
standard general relativity in 7 dimensions. This is due to the fact that all the
higher order terms of Lovelock gravity in 7 dimensions do not
contribute in the field equations.
The 7-dimensional solution given by Eqs. (\ref{metric}) and (\ref{frs}) presents a
black hole solution provided $f(r)$ has at least one real
positive root $r_{+}$. The existence of extreme black holes
depend on the existence of positive real root(s) for equation
$T=0$, which reduces to:
\begin{equation}
r_{\mathrm{m}}(3r_{\mathrm{m}}^{4}+2kl^{2}r_{\mathrm{m}}^{2}+al^{4})=0.
\label{rc6}
\end{equation}
The above equation shows that extreme black holes may exist only for
the case of $k=-1$.

Depending on the choice of the parameter $a$, the metric function $%
f(r)$ may have two minimums or one. Then, we have different
situations corresponding to different values of $a$:

\noindent \textbf{I.} For $a<1/3$, the metric function $f(r)$ has
two minimums located at the smallest and largest positive real
roots of Eq. (\ref{rc6}) denoted by
$r_{<}$ and $r_{>}$, respectively. In 7 dimensions, $r_{<}=0$, and $%
r_{>}=l\{1+\sqrt{1-3a}\}/\sqrt{3}\}^{1/2}$. The two minimums of the metric function $%
f(r)$ have the same value equal to zero for the special choice of
$a_{c}=1/4$, which is the solution of the following set of
equations:
\begin{equation}
f(r)|_{r=r_{<}}=0;\hspace{.5cm}f(r)|_{r=r_{>}}=0.  \label{dfrc}
\end{equation}
If $a\leq
a_{c}$, the value of $f(r_{<})\geq f(r_{>})$, while for $%
a> a_{c}$, $f(r_{<})<f(r_{>})$. Thus, we have two possibilities as follows:
\begin{figure}[ht]
\centering {\includegraphics[width=7cm]{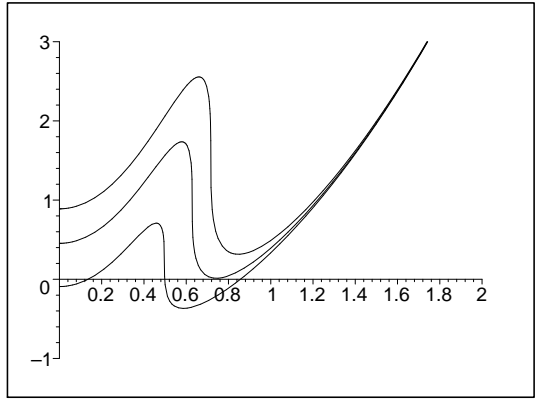} }
\caption{$f(r)$ vs. $r$ for $k=-1$, $n=6$, $a=0.2<a_c$, and
$m<m_{\mathrm{ext}}$, $m=m_{\mathrm{ext}}$ and
$m>m_{\mathrm{ext}}$ from up to down, respectively.} \label{Fr7a}
\end{figure}
\begin{figure}[ht]
\centering {\includegraphics[width=7cm]{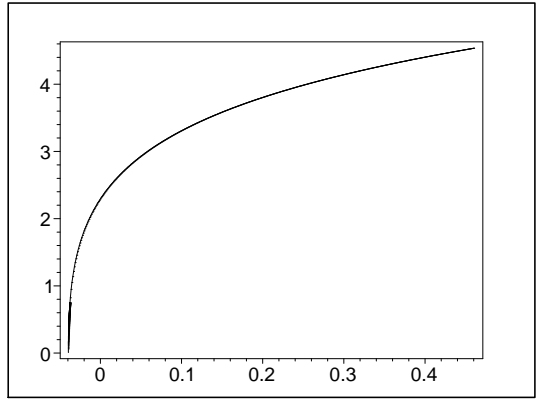}} \caption{$T$
vs. $m$ for $k=-1$, $n=6$, $a=0.2<a_c$.} \label{Tem7a}
\end{figure}

(i) If $a\leq a_c$, then $f(r_{<})\geq f(r_{>})$ and therefore
the mass of the extreme black hole may be obtained by computing Eq.
(\ref{bmass}) at $r_h=r_>$, i.e. $m_{\mathrm{ext}}=m(r_>)$. Then, the solution
given by Eqs. (\ref {metric}) and (\ref{frs}) presents a black
hole with inner and outer
horizons provided $m>m_{\mathrm{ext}}$, an extreme black hole if $m=m_{%
\mathrm{ext}}$, and a naked singularity for $m<m_{\mathrm{ext}}$
(see Fig. \ref{Fr7a}). Examining the local stability for $m\geq
m_{\mathrm{ext}}$ shows that the temperature versus $m$
monotonically increases from zero to infinity, as one may see in
Fig. \ref{Tem7a}. Thus, a hyperbolic black hole is locally stable,
if $a\leq a_{c}$.

\begin{figure}[ht]
\centering {\includegraphics[width=7cm]{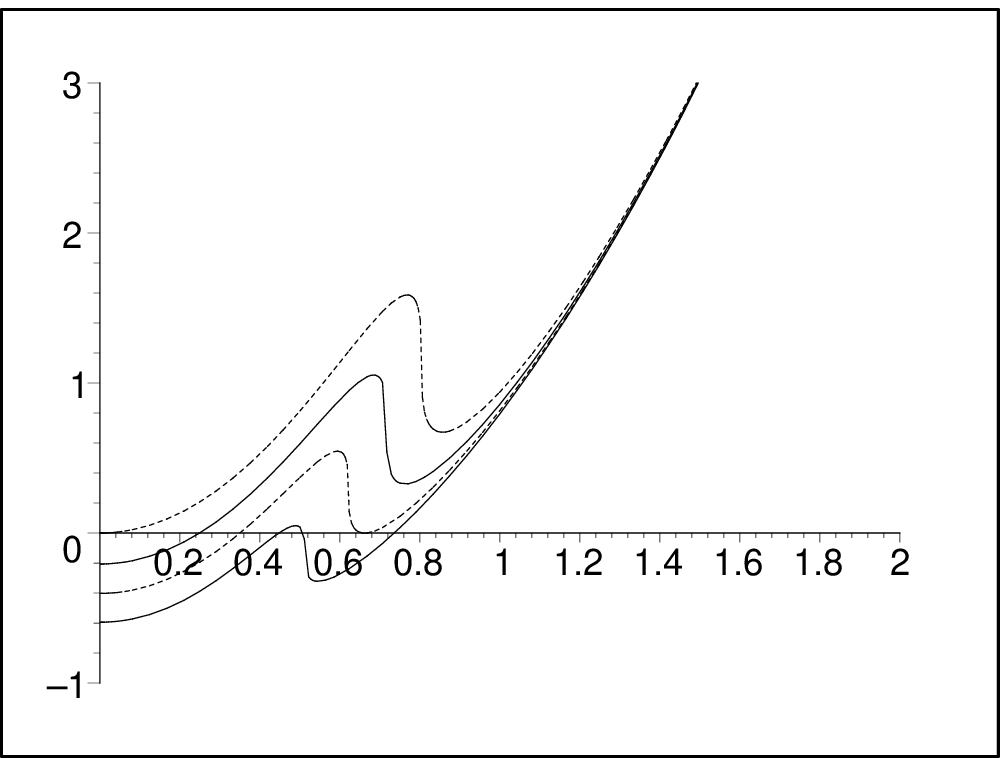} }
\caption{$f(r)$ vs. $r$ for $k=-1$, $n=6$, $a=0.3>a_c$, and
$m<m_{\mathrm{1ext}}$, $m=m_{\mathrm{1ext}}$,
$m_{\mathrm{1ext}}<m<m_{\mathrm{2ext}}$ ,$m=m_{\mathrm{2ext}}$,
and $m>m_{\mathrm{2ext}}$ from up to down, respectively.}
\label{Fr7b}
\end{figure}

\begin{figure}[ht]
\centering {\includegraphics[width=7cm]{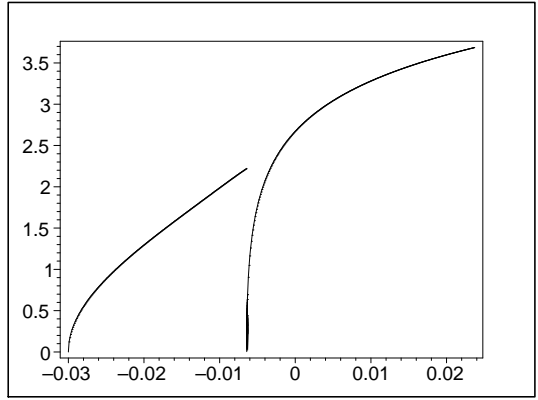}} \caption{$T$
vs. $m$ for $k=-1$, $n=6$, $a=0.3>a_c$.} \label{Tem7b}
\end{figure}

(ii) If $a_c<a<1/3$, then $f(r_{<})<f(r_{>})$. In this case, there
exist two values for the mass of the extreme black hole
$m_{\mathrm{1ext}}=m(r_<)$ and $m_{\mathrm{2ext}}=m(r_>)$. For
black holes with $m<m_{\mathrm{2ext}}$, the mass of the extreme
black hole is $m_{\mathrm{1ext}}$, while for black holes with
$m>m_{\mathrm{2ext}}$, the mass of the extreme black hole is
$m_{\mathrm{2ext}}$. Then, our solution presents a black hole
solution with event horizon radius, $r_{<}\leq r_{+}< r_{>}$ or
$r_+\geq r_{>}$ provided the mass parameter is in the range
$m_{\mathrm{1ext}}\leq m <m_{\mathrm{2ext}}$ or $m\geq
m_{\mathrm{2ext}}$, respectively. The metric function of these
black holes are shown in Fig. \ref{Fr7b}. The temperature versus
$m$ is shown in Fig. \ref{Tem7b}, which shows that the slope of the temperature versus
$m$ is always positive, and therefore these black holes are thermodynamically stable.
One may note that there is a discontinuity in this curve,
which is due to the fact that as $m$ approaches $m_{\mathrm{2ext}}$,
the radius of the event horizon suddenly
changes from $r_+<r_>$ to $r_+=r_>$.


\begin{figure}[ht]
\centering {\includegraphics[width=7cm]{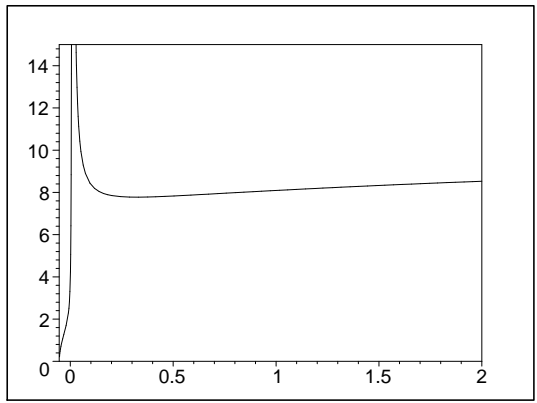}} \caption{$T$ vs.
$m$ for $k=-1$, $n=6$, $a=0.4$.} \label{Tem7c}
\end{figure}

\noindent \textbf{II.} For $a>1/3$, the metric function $f(r)$ has
just one minimum at
$r_{m}=r_{\mathrm{ext}}=0$, and the mass of the extreme black hole is $m_{\mathrm{ext%
}}=-a^2 l^4/3$ which is negative. In this case, we distinguish a
mass parameter $m_{\infty }=(3a-1)a^{2}l^{4}/3$, for which the
temperature becomes infinity. The plot of the temperature versus
$m$ (Fig. \ref{Tem7c}) shows that the temperature starts
from zero for $m=m_{\mathrm{ext}}$, goes to infinity as $m$ approaches $%
m_{\infty }$, decreases to a minimum and then increases. Thus, one
encounters with a Hawking-Page phase transition. This is a
peculiar feature of third order Lovelock gravity, that does not
occur in Einstein gravity \cite{birmingham} or Gauss-Bonnet gravity \cite{Cai1}.

The solution for a=1/3 presents a black hole of dimensionally continued Lovelock gravity in 7
dimensions with one horizon. In this case, the solution presents
a black hole provided $m\geq -l^4/27$, and the temperature monotonically increases from zero
to infinity. Thus, its slope is always positive and the black
hole is thermodynamically stable.

\subsection{$(n+1)$-dimensional hyperbolic black holes}

\begin{figure}[ht]
\centering {\includegraphics[width=7cm]{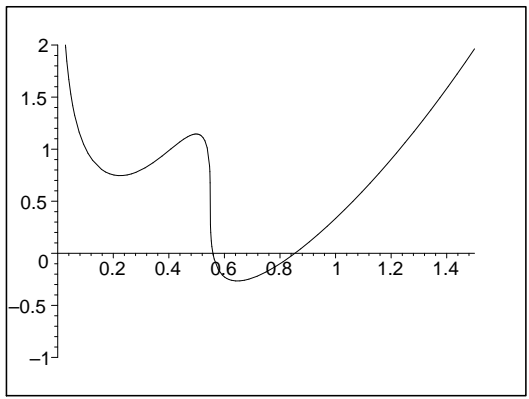}} \caption{$f(r)$
vs. $r$ for $k=-1$, $n=8$.} \label{Fr8}
\end{figure}

We can easily extend all of the discussions of the previous subsection to ($n+1$%
)-dimensional solutions. The extrema of the metric function
$f(r)$ are located at the roots of the following equation:
\begin{equation}
3nr_{m}^{6}-(3n-6)l^{2}r_{m}^{4}+(3n-12)al^{4}r_{m}^{2}-(n-6)a^{2}l^{6}=0.
\label{rcn}
\end{equation}
Depending on the choice of the parameter $a$, as in seven dimensions, $%
f(r)$ might have two minimums or one. Indeed, $f(r)$ has two
minimums provided $a^{(-)}\leq a<a^{(+)}$, where
\begin{equation}
a^{(-)}=\frac{(n-8)(n-2)^{2}}{3n(n-6)^{2}},\qquad
a^{(+)}=\frac{1}{3},
\end{equation}
and has one minimum otherwise. Of course, one may note that
$a^{(-)}$ is only positive for $n>8$. Thus, we discuss the following three cases separately:

\noindent \textbf{I.} For $a^{(-)}\leq a<a^{(+)}$, the metric function $%
f(r)$ has two minimums located at $r_{<}$ and $r_{>}$ related to
the smallest and largest positive real roots of (\ref{rcn}). It
is worth to mention that the smallest positive real root of
(\ref{rcn}) is not zero. As in the case of part (\textbf{I}) of
the previous subsection, there exists a critical value for the
parameter $a_{c}$, for which the set of equations (\ref{dfrc})
hold, and the value of $f(r_{<})=f(r_{>})$ for $a=a_{c}$. The values of $%
a_{c}$ are $0.3018,0.3169,0.3237,0.3271$ for
$n=7,8,9,10$, respectively. The diagram of $f(r)$ is slightly
different from it's analogous seven-dimensional case, as one may
see in Fig. (\ref{Fr8}). With these modifications, all of
our discussions in (\textbf{I}) are still valid in this case, and the
black holes are stable.

\noindent \textbf{II.} For $a\geq a^{(+)}$, there is only one real root for (\ref{rcn}%
), which means that $f(r)$ has just one minimum at
$r_{\mathrm{ext}}$. Here again, all the conclusions are similar
to the case of 7-dimensional solutions with $a>1/3$ discussed in
the previous subsection. That is, one encounters with an unstable
phase for the black hole.

\noindent \textbf{III.} For $n>8$, there is a region $0<a<a^{(-)}$,
for which $f(r)$ has just one minimum at $r_{\mathrm{ext}}$
corresponding to the only
positive real root of (\ref{rcn}). We have black hole interpretation for $%
m>m(r_{\mathrm{ext}})$ and the temperature is always monotonically
increasing. Thus, the black hole solutions are always stable in
this region.
\subsection{Spherical black holes}
\begin{figure}[ht]
\centering {\includegraphics[width=7cm]{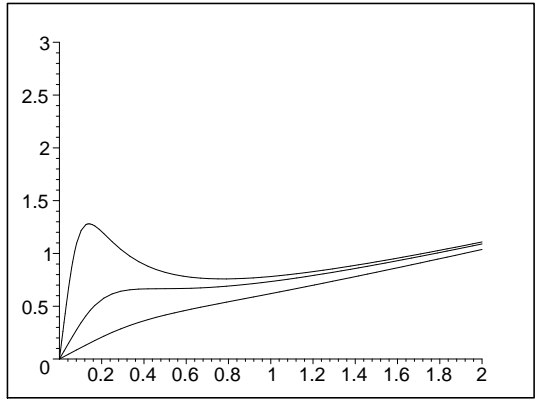} } \caption{$T$
vs. $r_+$ with $k=1$ and $n=6$ for $a<a_c$, $a=a_c=.046$ and
$a>a_c$ from up to down, respectively.} \label{Tempk17}
\end{figure}

In the case of $k=1$, the solution presents a black hole with one
horizon at $r_+$ provided the mass of it is greater than a critical value $m_c$.
For these black holes, the temperature is always
positive and there is no extreme black hole. To analyze the stability,
one may plot the temperature versus $r_+$. The plot of temperature versus $r_+$
for a 7-dimensional black hole is shown in Fig. \ref{Tempk17}. This figure shows that
for small values of Lovelock coefficient, there exist
an intermediate unstable phase, while for large values of Lovelock coefficients
the black hole is stable.

\begin{figure}[ht]
\centering {\includegraphics[width=7cm]{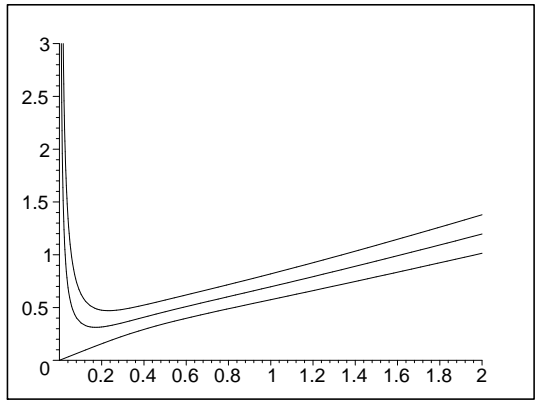} } \caption{$T$
vs. $r_+$ with $k=1$ and $a=0.2$ for $n+1=7, 8$ and $10$ from down
to up, respectively.} \label{Tempk1HD}
\end{figure}

But in higher dimensions, there exists only an intermediate
unstable phase for all values of Lovelock coefficient as one may see
in Fig. (\ref{Tempk1HD}).

In comparison with the asymptotically AdS spherical black holes of Einstein gravity,
which have a small unstable phase, the stability phase structure of
the black holes of third order Lovelock gravity with spherical horizon shows that the Lovelock term
changes the stability phase structure.

\section{Concluding Remarks}
The topological black holes with hyperbolic horizon in Einstein and second order Lovelock gravities are
stable \cite{birmingham,Cai1}. Also, the Lovelock terms do not change the
stability phase structure of a black hole with flat horizons \cite{deh}. These facts bring
in the idea that the stability of topological black holes may be a robust
feature of Lovelock gravity. In this paper, we studied the phase structure of topological black holes
of third order Lovelock gravity with hyperbolic horizons, and found that they
have an intermediate unstable phase for large values of third order Lovelock coefficient.
That is, when the effect of third order Lovelock term becomes
more relevant, then an unstable phase start appearing. This drastic change in the
stability of topological black holes of third order Lovelock gravity persuaded us
to investigate the effects of third order Lovelock term on the stability of black holes with
spherical horizon.
We found that a 7-dimensional spherical black hole in third
order Lovelock gravity has an intermediate unstable phase for small
third order Lovelock coefficient and is stable for large $\alpha_3$.
That is, the third order Lovelock term changes the stability behavior
of a black hole, but this effect is not peculiar to third order Lovelock
gravity and it occurs in Gauss-Bonnet gravity too \cite{Cai1}.
It is worth to mention that an asymptotically AdS black hole in Einstein gravity
with small mass is thermodynamically unstable, while in Lovelock gravity
it is stable for large values of Lovelock coefficients.
This stability analysis shows that the stability of black
holes with curved horizons
is not a robust feature of all the generalized theory of gravity. Also, we found that
the entropy of third order Lovelock gravity reduces to the area law of
the entropy for $\alpha_2=\alpha_3=0$. But, as in the case
of Einstein gravity, it does not go to zero
for the extreme black holes whose temperature is zero.

Although the topological black holes of third order Lovelock gravity are
thermodynamically unstable for large values of third order Lovelock coefficient,
it is worth to examine the dynamical (gravitational)
instability of these black holes. This is due to the fact that there are black holes
in Einstein gravity which are thermodynamically unstable, while they are dynamically stable \cite{Kon}.
However, there may be some correlations between the dynamic and thermodynamic
instability of black hole solutions of other generalized theory of gravity \cite{Mhessian}.
\acknowledgments
This work has been supported by Research Institute for Astrophysics and
Astronomy of Maragha.

\end{document}